\documentclass[aps,prl,twocolumn]{revtex4}%
\usepackage{amsfonts}
\usepackage{version}
\usepackage{amsmath}
\usepackage{amssymb}
\usepackage{graphicx}
\providecommand{\U}[1]{\protect\rule{.1in}{.1in}}

\ifx\pdfoutput\relax\let\pdfoutput=\undefined\fi
\newcount\msipdfoutput
\ifx\pdfoutput\undefined\else
\ifcase\pdfoutput\else
\ifx\paperwidth\undefined\else
\ifdim\paperheight=0pt\relax\else\pdfpageheight\paperheight\fi
\ifdim\paperwidth=0pt\relax\else\pdfpagewidth\paperwidth\fi
\fi\fi\fi
\begin{document}
\title{Coherence of an optically illuminated single nuclear spin qubit}
\author{L. Jiang,$^{1}$ M. V. Gurudev Dutt,$^{1}$ E. Togan,$^{1}$ L. Childress,$^{1}$ P.
Cappellaro,$^{2}$ J. M. Taylor,$^{3}$ M. D. Lukin$^{1,2}$}
\affiliation{$^{1}$ Department of Physics, Harvard University, Cambridge, MA 02138}
\affiliation{$^{2}$ Institute for Theoretical Atomic, Molecular and Optical Physics,
Cambridge, MA 02138}
\affiliation{$^{3}$ Department of Physics, Massachusetts Institute of Technology,
Cambridge, MA 02139}


\begin{abstract}
We investigate the coherence properties of individual nuclear spin quantum
bits in diamond [\mbox{Dutt \emph{et al.},} \mbox{Science}, \textbf{316}, 1312
(2007)] when a proximal electronic spin associated with a nitrogen-vacancy
(NV) center is being interrogated by optical radiation. The resulting nuclear
spin dynamics are governed by time-dependent hyperfine interaction associated
with rapid electronic transitions, which can be described by a spin-fluctuator
model. We show that due to a process analogous to motional averaging in
nuclear magnetic resonance, the nuclear spin coherence can be preserved after
a large number of optical excitation cycles. Our theoretical analysis is in
good agreement with experimental results. It indicates a novel approach that
could potentially isolate the nuclear spin system completely from the
electronic environment.

\end{abstract}
\date{\today}
\maketitle

Nuclear spins are of fundamental importance for storage and processing of
quantum information. Their excellent coherence properties make them a superior
qubit candidate even in room temperature solids. Unfortunately, their weak
coupling to the environment also makes it difficult to isolate and manipulate
individual nuclei. Recently, coherent preparation, manipulation and readout of
individual $^{13}$C nuclear spins in the diamond lattice were demonstrated
\cite{Dutt06, Jelezko04}. These experiments make use of optical polarization
and manipulation of the electronic spin associated with a nitrogen-vacancy
(NV) color center in the diamond lattice
\cite{Awschalom07,Childress06,Gaebel06,Hanson06}. This enables reliable
control of the nuclear spin qubit via hyperfine interactions with the
electronic spin.

In order to be useful for applications in scalable quantum information
processing \cite{Awschalom07}, such as quantum communication
\cite{Childress05} and quantum computation \cite{JTSL06}, the quantum
coherence of the nuclear spins must be maintained even when the electronic
state is undergoing fast transitions associated with optical measurement and
with entanglement generation between electronic spins. In this Letter, we
investigate coherence properties of such an optically illuminated nuclear
spin--electron spin system. We show that these properties are well-described
by a spin-fluctuator model \cite{Paladino02,Galperin03,Galperin06,Bergli06},
involving a single nuclear spin (system) coupled by the hyperfine interaction
to an electron \cite{FootNote1} (fluctuator) that undergoes rapid optical
transitions and mediates the coupling between the nuclear spin and the
environment. We generalize the spin-fluctuator model to a vector description,
necessary for single NV centers in diamond \cite{Dutt06}, and make direct
comparisons with experiments. Most importantly we demonstrate that the
decoherence of the nuclear spin due to the rapidly fluctuating electron is
greatly suppressed via a mechanism analogous to motional narrowing in nuclear
magnetic resonance (NMR) \cite{Slichter90,Happer77}, allowing the nuclear spin
coherence to be preserved even after hundreds of optical excitation cycles. We
further show that by proper tuning of experimental parameters it may be
possible to completely decouple the nuclear spin system from the electronic
environment.
The spin-fluctuator model discussed here for NV centers can be generalized to
other AMO systems (see Refs. \cite{Boyd06, Reichenbach07} for the recent
progress).%

\begin{figure}[t]
\begin{center}
\includegraphics[
width=8.5cm
]{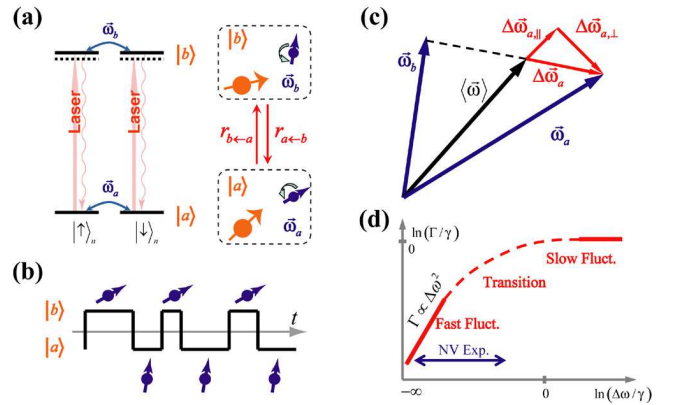}
\end{center}
\caption[figure1]{(a) Energy levels (left) and schematic model
(right) for optical transitions between different electronic states
($\left\vert a\right\rangle $ and $\left\vert b\right\rangle $), with
transition rates $r_{ba}$ and $r_{ab}$. The precession of the nuclear spin
($\vec{\omega}_{a}$ or $\vec{\omega}_{b}$) (blue arrow) depends on the
electronic state ($\left\vert a\right\rangle $ or $\left\vert b\right\rangle
$) (orange arrow). (b) Random telegraph trajectory of the
electron, and time-dependent precession of the nuclear spin. (c) Geometric
representation of Larmor precession vectors. (d) The decoherence rate $\Gamma$
as a function of the differential precession frequency $\Delta\omega$, in
units of $\gamma$.}%
\label{figure1}%
\end{figure}

The essential idea of this work is illustrated in Fig.~\ref{figure1}. We
consider an individual nuclear spin system ($I=1/2$, associated with a $^{13}%
$C atom) that is weakly coupled to the electronic spin of an NV center via the
hyperfine interaction. The transitions between ground and optically excited
electronic states are caused by laser light and spontaneous emission of
photons. The strength of the hyperfine interaction differs between the ground
and the excited electronic states, because they have different spatial wave
functions and thus different overlap with the nucleus. As the electron
undergoes rapid optical transitions, the nuclear spin precesses according to a
time-dependent effective magnetic field induced by the electron.

This situation can be modeled by considering the electron as a fluctuator with
two states, $\left\vert a\right\rangle $ and $\left\vert b\right\rangle $. Let
us first assume that the incoherent transitions between these two electronic
states $\left\vert a\right\rangle \overset{r_{ba}}{\underset{r_{ab}%
}{\rightleftharpoons}}\left\vert b\right\rangle $ are described by the random
telegraph process as shown in Fig.~\ref{figure1}(b), which is fully
characterized by the classical transition rates $r_{ba}$ and $r_{ab}$
(corresponding to the optical pumping rate and the radiative decay rate,
respectively, resulting from an off-resonant optical drive). The nuclear spin
will undergo time-dependent precession, characterized by vectors $\vec{\omega
}_{a}$ and $\vec{\omega}_{b}$ for the electronic states $\left\vert
a\right\rangle $ and $\left\vert b\right\rangle $, respectively
[Fig.~\ref{figure1}(a)].

In general, the precession vectors $\vec{\omega}_{a}$ and $\vec{\omega}_{b}$
may point along different directions as shown in Fig.~\ref{figure1}(c). For
example, the nuclear spin can precess around different axes for different
electronic states. Furthermore, the electron undergoes fast optical
transitions and introduces high frequency noise, which, in addition to
dephasing, can induce spin-flips \cite{Dutt06}. Therefore, we need to consider
the nuclear spin precession around a time-dependent, stochastic vector
$\vec{\omega}\left(  t\right)  $, generalizing the earlier scalar model
\cite{Galperin06,Bergli06}.

Let $\gamma$ and $\Delta\omega$ be the typical scales for the electron
transition rates and the \emph{difference} between the qubit precession
vectors, respectively. Let us now consider the limiting case of a
fast-fluctuator ($\gamma\gg\Delta\omega$). In this case we may use a
perturbative approach associated with the small phase shift acquired by the
nuclear spin during one excitation cycle $\delta\phi\sim\Delta\omega/\gamma$.
On average this phase shift will result in a modification of the precession
frequency. In addition, due to random variations in the time spent in
different electronic states the phase shift will undergo a random walk with
diffusion constant $\sim\delta\phi^{2}\times\gamma\sim\Delta\omega^{2}/\gamma$.

More precisely, to the first order, we have the average precession vector%
\begin{equation}
\left\langle \vec{\omega}\right\rangle =\frac{r_{ba}^{-1}\vec{\omega}%
_{a}+r_{ab}^{-1}\vec{\omega}_{b}}{r_{ba}^{-1}+r_{ab}^{-1}}, \label{LarmorAvg}%
\end{equation}
where the weights are proportional to the durations of different states for
the fluctuator. As illustrated in Fig.~\ref{figure1}(c), $\left\langle
\vec{\omega}\right\rangle \ $defines the quantization axis of the spin system.
The difference between the instantaneous precession vector ($\vec{\omega}_{a}$
or $\vec{\omega}_{b}$) and $\left\langle \vec{\omega}\right\rangle $,
$\Delta{\vec{\omega}}=\vec{\omega}_{a}-\left\langle \vec{\omega}\right\rangle
$ can be decomposed into the parallel and perpendicular components. The
perpendicular component introduces spin-flips along the quantization axis at
rate $\Gamma_{1}\sim\left(  \Delta\omega\right)  _{\perp}^{2}/\gamma$. The
parallel component causes stochastic phase accumulation, leading to dephasing
at the rate $\Gamma_{\phi}\sim\left(  \Delta\omega\right)  _{\parallel}%
^{2}/\gamma$. Note that both rates are inversely proportional to the
fluctuator transition rate $\gamma$ in the limit of fast electronic
transitions. The underlying physics is analogous to the motional narrowing of
NMR \cite{Slichter90}, in which the rapid motion of the environment
(corresponding to large $\gamma$) averages out the randomly accumulated phase.

In the opposite slow-fluctuator limit ($\gamma\lesssim\Delta\omega$), the
decoherence rate is only determined by the fluctuator transition rates. For
each fluctuator transition, there is a time variation, $\delta t\sim1/\gamma$,
which induces an uncertainty in the rotation $\delta\phi\sim\Delta\omega\delta
t\sim\Delta\omega/\gamma\sim1$. This implies that a single transition of the
fluctuator is sufficient to destroy the coherence of the spin system. The
resulting qualitative dependence of the nuclear spin decay upon difference in
Larmor precession is illustrated in Fig.~\ref{figure1}(d).

We now introduce the master equation formalism and illustrate that it is
possible to reduce the system dynamics to a set of linear differential
equations, even in the presence of the non-commutative stochastic precession.
We will first solve the spin-fluctuator model with the two-state fluctuator
described above. After that, we extend the procedure to include multi-state
fluctuators into the formalism.

The incoherent transition of the two-state fluctuator [Fig.~\ref{figure1}(a)]
can be described by the master equations%
\begin{equation}
\frac{d}{dt}\left(
\begin{array}
[c]{c}%
p_{a}\\
p_{b}%
\end{array}
\right)  =\left(
\begin{array}
[c]{cc}%
-r_{ba} & r_{ab}\\
r_{ba} & -r_{ab}%
\end{array}
\right)  \left(
\begin{array}
[c]{c}%
p_{a}\\
p_{b}%
\end{array}
\right)  , \label{ME1}%
\end{equation}
where $p_{a}$ and $p_{b}$ are occupation probabilities for states $\left\vert
a\right\rangle $ and $\left\vert b\right\rangle $.

The Hamiltonian of the nuclear spin (depending on the state of the fluctuator)
is $H=\left\vert a\right\rangle \left\langle a\right\vert \otimes
H_{a}+\left\vert b\right\rangle \left\langle b\right\vert \otimes H_{b}$, with
$H_{a}=\vec{\omega}_{a}\cdot\vec{I}$ and $H_{b}=\vec{\omega}_{b}\cdot\vec{I}$.

Since there is no coherence between different fluctuator states, we may write
the density matrix for the composite system as $\rho=\left\vert a\right\rangle
\left\langle a\right\vert \otimes\rho_{a}+\left\vert b\right\rangle
\left\langle b\right\vert \otimes\rho_{b}$, where $\rho_{j}=\left(
\begin{array}
[c]{cc}%
\rho_{j,11} & \rho_{j,12}\\
\rho_{j,21} & \rho_{j,22}%
\end{array}
\right)  $ for $j=a$ or $b$. The unitary evolution of the density matrix
$\rho$ with Hamiltonian $H$ can be decomposed into two uncoupled parts:
$\frac{d}{dt}\rho_{j}=-i\left[  H_{j},\rho_{j}\right]  $ for $j=a,b$. In terms
of the Liouville operator, the unitary evolution is%
\begin{equation}
\frac{d}{dt}\vec{\rho}_{j}=\mathcal{L}_{j}\vec{\rho}_{j},
\end{equation}
where the density operator is represented by a column vector $\vec{\rho}%
_{j}=\left(  \rho_{j,11},\rho_{j,12},\rho_{j,21},\rho_{j,22}\right)  ^{T}$ and
the Liouville operator by a matrix%
\begin{equation}
\mathcal{L}_{j}\equiv\mathcal{L}\left[  \vec{\omega}_{j}\right]  \equiv\left(
-i\right)  \left(  H_{j}\otimes\mathbf{I}-\mathbf{I}\otimes H_{j}^{\ast
}\right)  ,
\end{equation}
for $j=a,b$. Notice that the transition matrix depends linearly on the
precession vector, and such linearity implies $\mathcal{L}\left[  \vec{\omega
}_{a}\right]  +\mathcal{L}\left[  \vec{\omega}_{b}\right]  =\mathcal{L}\left[
\vec{\omega}_{a}+\vec{\omega}_{b}\right]  $.

Combining the dynamics of the system and the fluctuator, we may write down the
following master equations:%
\begin{equation}
\frac{d}{dt}\left(
\begin{array}
[c]{c}%
\vec{\rho}_{a}\\
\vec{\rho}_{b}%
\end{array}
\right)  =\left(
\begin{array}
[c]{cc}%
\mathcal{L}_{a}-\mathbf{r_{ba}} & \mathbf{r_{ab}}\\
\mathbf{r_{ba}} & \mathcal{L}_{b}-\mathbf{r_{ab}}%
\end{array}
\right)  \left(
\begin{array}
[c]{c}%
\vec{\rho}_{a}\\
\vec{\rho}_{b}%
\end{array}
\right)  , \label{ME3}%
\end{equation}
where $\mathcal{L}_{a}$ and $\mathcal{L}_{b}$ describe the (slow) dynamics of
the precession; $\mathbf{r_{ba}}=r_{ba}~\mathbf{I}_{4\times4}$ and
$\mathbf{r}_{\mathbf{ab}}=r_{ab}~\mathbf{I}_{4\times4}$ are associated with
the (fast) incoherent optical transitions between electronic states, not
affecting the nuclear spin.

We generalize to a multi-state fluctuator, by introducing $r_{ij}$ the
fluctuator transition rate from state $j$ to state $i$, and $r_{jj}\equiv
\sum_{i\neq j}r_{ij}$ the total transition rate from state $j$ to all other
states. For an $N$-state fluctuator, the generalized form for Eq.(\ref{ME3})
is%
\begin{equation}
\frac{d}{dt}\vec{\rho}_{i}=\left(  \lambda\mathcal{L}_{i}-\mathbf{r}%
_{ii}\right)  \vec{\rho}_{i}+\sum_{j\neq i}^{N}\mathbf{r}_{ij}\vec{\rho}_{j}
\label{ME general}%
\end{equation}
where $\vec{\rho}_{j}=\left(  \rho_{j,11},\rho_{j,12},\rho_{j,21},\rho
_{j,22}\right)  ^{T}$ for $j=1,2,\cdots,N$, and $\mathbf{r}_{ij}%
=r_{ij}\mathbf{I}_{4\times4}$. When there are $M$ fluctuators, with $N_{j}$
states for the $j$th fluctuator, we can always reduce it to one composite
fluctuator with $N=\prod_{j=1}^{M}N_{j}$ states.

Given all the parameters $\left\{  \vec{\omega}_{i}\right\}  $ and $\left\{
r_{ij}\right\}  $, we can solve $\vec{\rho}_{i}\left(  t\right)  $ exactly
from the above $4N$ linear differential equations [Eq.(\ref{ME general})] with
initial conditions for $\left\{  \vec{\rho}_{i}\left(  0\right)  \right\}  $.
Finally, the density matrix of the nuclear spin system is $\vec{\rho}\left(
t\right)  =\sum_{i}\vec{\rho}_{i}\left(  t\right)  $, which together with Eq.
(\ref{ME general}) provides an exact solution to the generalized spin-fluctuator model.

The experimental procedure for probing the dynamics of an optically
illuminated nuclear spin qubit proximal to NV centers in diamond is described
in detail in Ref.~\cite{Dutt06}. The NV center is a spin triplet in the ground
electronic state. In the experiment we optically polarize the electron into
$m_{s}=0$, in which the hyperfine interaction with the nuclear spin vanishes
to leading order. Furthermore, it is believed \cite{Hemmer01}, and is
confirmed by experimental evidence reported below, that the electronic spin is
a good quantum number during the optical excitation of the NV center. Hence
the electron should mostly remain in the $m_{s}=0$ manifold during optical excitation.

\begin{figure}[b]
\begin{center}
\includegraphics[
width=8.0cm
]{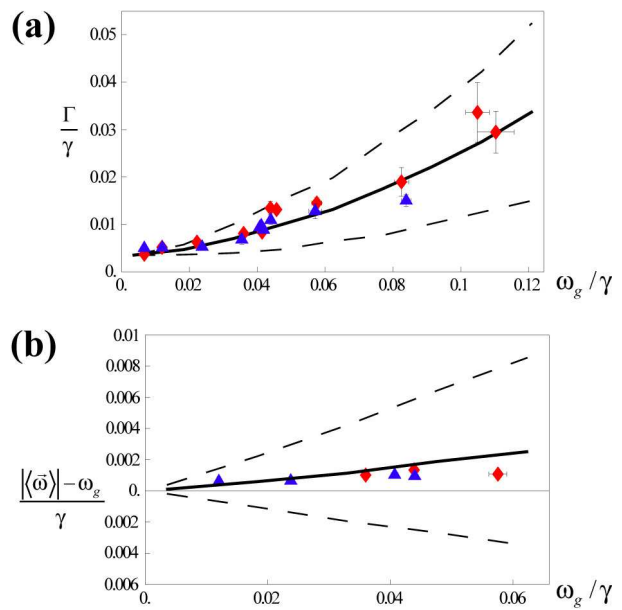}
\end{center}
\caption[figure2]{Comparison between experimental data (points) and simulation
(lines). (a) Optically induced decoherence rate $\Gamma$ as a function of
ground state Larmor precession frequency $\omega_{g}$ (data adopted
from Ref. \cite{Dutt06}). (b) Shift of average Larmor precession
frequency $\left\vert \left\langle \vec{\omega}\right\rangle \right\vert
-\omega_{g}$ as a function of $\omega_{g}$. For both plots, the axes are also
labeled in dimensionless units, normalized by the radiative decay rate
$\gamma=86$ $\mu$s$^{-1}$. Experimental data points (blue triangles, red
diamonds) include nuclear spins prepared along both directions ($\hat{x}$,
$\hat{z}$) perpendicular to the Larmor precession vector ($\hat{y}$). The full
lines are from simulation of the generalized spin-fluctuator model, averaging
over 50 different sets of randomly generated excited states, as described in
the text. The simulation assumes $N=3$ for the fluctuator (i.e. one ground
state and two excited states \cite{Manson06,Lenef96}). The dashed lines are
the statistical standard deviation of the different realizations. In panel
(a), the curves from simulation are manually shifted upwards by $\Gamma
_{0}=3.4\times10^{-3}\gamma$, as described in the text.}%
\label{figure2}%
\end{figure}

However, the hyperfine interaction with the electron can dramatically change
the precession of the nuclear spin by modifying its effective magnetic
moment~\cite{Childress06}. The direction and magnitude of the precession
vector, which is determined by the effective $g$-tensor \cite{Childress06},
varies due to the changes in the contact and dipolar interactions associated
with different electronic states. Under these experimental conditions, the
nuclear precession vectors associated with different electronic states should
be proportional to the perpendicular components of the external magnetic
field, $B_{\perp}$, with a proportionality constant and direction that depends
upon the electronic state. Thus, we present the experimental data
(Fig.~\ref{figure2}) as functions of the ground state precession frequency
$\omega_{g}$ ($\omega_{g}\propto B_{\perp}$), which can be accurately
measured. Both the optically induced decoherence rate $\Gamma$ (the decay rate
of the nuclear spin Bloch vector) and the change in average Larmor precession
frequency $\left\langle \vec{\omega}\right\rangle -\omega_{g}$ were measured
for a particular NV center. For the presented data, the optical excitation
rate was chosen to correspond to about one half of saturation
intensity.

A comparison between first-principle theory and experiment would require
precise knowledge of nuclear precession vectors for both the \textit{ground
}and\textit{ excited} electronic states. The experiments, performed at room
temperature, involve excitation of multiple excited states, whose wave
functions are not known in great detail. To model quantum dynamics of such a
system, we assume that the excited state precession vector has similar order
of magnitude to that of the ground state, but might point along a different
direction. In the following, we label the ground state as the first state of
the fluctuator with precession frequency $\vec{\omega}_{g}\equiv\vec{\omega
}_{1}$ for the proximal nuclear spin. The $j$th excited state has precession
vector $\vec{\omega}_{j}$, with its three components drawn from a normal
distribution with mean $0$ and deviation $\sigma_{\omega}\sim\omega_{g}$, for
$j=2,\cdots,N$. We assume that the excitation rate from the $1$st to the $j$th
excited state $r_{j1}=R/\left(  N-1\right)  $, the radiative decay rate
$r_{1j}=\gamma=86$~$\mu$s$^{-1}$ \cite{Manson06}, and the total excitation
rate $R=\gamma$. The transitions among excited states are neglected. According
to \cite{Manson06,Lenef96}, there are at least two excited states involved in
the optical transition, so we set $N=3$. By choosing $\sigma_{\omega
}=2.5\omega_{g}$, we find good agreement between theory and experiment as
shown in Fig.~\ref{figure2}.

In the fast-fluctuating regime ($\omega_{g}\ll\gamma$), the experimental
decoherence rate increases quadratically with $\omega_{g}$, consistent with
the scaling obtained from the fast-fluctuator limit. When the precession
frequency becomes comparable to the fluctuator transition rates ($\omega
_{g}\lesssim0.2\gamma$), $\Gamma$ increases sub-quadratically with $\omega
_{g}$. This is because we are in the transition region as indicated in
Fig.~\ref{figure1}(d). In principle, for even higher precession frequency, the
decoherence rate should saturate at the optical transition rate.
Experimentally, however, it is difficult to manipulate the nuclear spin for
very high precession frequency ($\omega_{g}>0.2\gamma$) \cite{Dutt06}.

In addition to the electronic spin-conserving optical transitions analyzed
above, the spin-changing transitions between $m_{s}=0$ and $m_{s}=\pm1$ may
also induce decoherence of the nuclear spin. However, the hyperfine field from
the electron in spin state $m_{s}=\pm1$ is oriented along the well-defined
$z$-axis \cite{Dutt06}, which introduces decoherence for nuclear spin state
$\left\vert \uparrow\right\rangle _{X}$, but not for $\left\vert
\uparrow\right\rangle _{Z}$. This contradicts the observation that the
decoherence rates (with initial states $\left\vert \uparrow\right\rangle _{X}$
and $\left\vert \uparrow\right\rangle _{Z}$) are similar for the observed
center (see Fig.~\ref{figure2}). Therefore, we conclude that the spin-changing
transitions should not be the dominant process for optically induced nuclear
spin decoherence.

By extrapolating the experimental data to zero external magnetic field, we
find that there is still a finite decoherence rate $\Gamma_{0}\approx
3.4\times10^{-3}\gamma$ (simulation curves are offset with additional fitting
parameter in Fig.~\ref{figure2}(a)). Remarkably, these data indicate that the
nuclear spin coherence is still maintained after scattering $\gamma/\Gamma
_{0}\sim300$ photons by the electron. This insensitivity, enabled via effects
analogous to motional-averaging, is of critical importance for the feasibility
of NV-center-based distant quantum communication \cite{Childress05} and
distributed quantum computation \cite{JTSL06} protocols.

The zero field decoherence rate $\Gamma_{0}$ is still related to optical
transitions, because it is much larger than the observed decoherence rate of
the nuclear spin in the dark $\Gamma_{dark}\approx3\times10^{-4}\gamma$
\cite{Dutt06}. We attribute this zero field decoherence to the orbital motion
of the optically excited states, which produces a \textquotedblleft
residual\textquotedblright\ magnetic field at the position of the nucleus. The
residual magnetic field can be present for optically excited states, because
the orbital motion for these states is not quenched \cite{Lenef96,Manson06}.
Considering the nucleus on the second coordination sphere with respect to the
vacancy (i.e., $r\approx2.6%
\operatorname{\text{\AA}}%
$), the magnetic field from the orbital motion is approximately $\mu_{B}%
/r^{3}\approx500\sim1000$ G. This gives an estimated decoherence rate
$\Gamma_{0}\ \approx\Delta\omega^{2}/\gamma\approx\left(  1\sim5\right)
\times10^{-3}~\gamma$, which is consistent with the value observed experimentally.

These observations may allow us to develop new methods to further suppress
optically induced nuclear decay. Specifically, at low temperatures ($T<10$ K),
it is possible to resolve individual optical transitions and selectively drive
the electron between the ground state and one excited state. Under these
conditions, it should be possible to eliminate the decoherence $\Gamma_{0}$ by
applying an external magnetic field that exactly compensates the residual
field from the orbital motion. With the compensation at this \textquotedblleft
sweet spot\textquotedblright, the nuclear spin sees the same total magnetic
field, regardless of the state of the electron, and therefore can be
completely decoupled from the electronic environment \footnote[100]
{The decoherence associated with the fluctuations of the compensating field
(e.g., $\delta B\sim0.1$ G and $\delta\omega\sim10^{-3}\Delta\omega$) is
approximately $\delta\omega^{2}/\gamma\sim10^{-6}\Gamma_{0}$, neglegible
compared with other decoherence processes.%
}
\footnote[200]
{Ionization of the NV center caused by green light might also lead to optically induced decoherence of the nuclear spin at B=0. Note that ionization probability should be reduced if resonant red light is used.}.

In conclusion, we have shown that the vector spin-fluctuator model provides a
good description for our observations of coherence properties of the optically
illuminated nuclear spin qubit. Our theoretical and experimental results
demonstrate a substantial suppression of nuclear spin decoherence due to the
mechanism analogous to the motional averaging in NMR. Our analysis further
indicated a new approach that may allow us to completely decouple the nuclear
spin and the electron during optical excitation. These results are of critical
importance for scalable applications of NV-center-based quantum registers
\cite{Childress05,JTSL06}.

We thank P. Hemmer, F. Jelezko and A.Zibrov for useful discussions and
experimental help. This work was supported by NSF (CAREER and PIF programs),
the ARO MURI, the Packard and Hertz Foundations.

\end{document}